\documentclass[12pt,preprint]{aastex}

\def\gtorder{\mathrel{\raise.3ex\hbox{$>$}\mkern-14mu
             \lower0.6ex\hbox{$\sim$}}}

\def\ltsima{$\; \buildrel < \over \sim \;$}
\def\simlt{\lower.5ex\hbox{\ltsima}}
\def\gtsima{$\; \buildrel > \over \sim \;$}
\def\simgt{\lower.5ex\hbox{\gtsima}}

\begin{document}


\title{On the progenitor of SN 2005gl and the nature of Type IIn supernovae}


\author{Avishay Gal-Yam\altaffilmark{1}}
\affil{Astronomy Department, MS 105-24, California Institute of Technology,
Pasadena, CA 91125}
\email{avishay@astro.caltech.edu}
\author{D. C. Leonard}
\affil{Department of Astronomy, San Diego State University, San Diego,
California 92182}
\author{D. B. Fox}
\affil{Department of Astronomy and Astrophysics,
Pennsylvania State University, 525 Davey Lab, University Park, PA 16802}
\author{S. B. Cenko, A. M. Soderberg, D.-S. Moon}
\affil{Division of Physics, Mathematics and Astronomy, California Institute of Technology,
Pasadena, CA 91125}
\author{D. J. Sand\altaffilmark{2}}
\affil{Steward Observatory, University of Arizona, 933 North Cherry Avenue,
Tucson, AZ 85721}
\author{(the CCCP)}
\and
\author{W. Li, A. V. Filippenko}
\affil{Department of Astronomy, 601 Campbell Hall, University of California, Berkeley, CA 94720-3411}
\author{G. Aldering}
\affil{E. O. Lawrence Berkeley National Laboratory, 1 Cyclotron Road, Berkeley, CA 94720}
\author{Y. Copin (for the SNfactory team)}
\affil{Institut de Physique Nucleaire de Lyon, France}

\altaffiltext{1}{Hubble Fellow.}
\altaffiltext{2}{Chandra Fellow.}


\begin{abstract}

We present a study of the type IIn supernova (SN) 2005gl, in the relatively
nearby ($d\approx66$ Mpc) galaxy NGC 266. Photometry and spectroscopy of the
SN indicate it is a typical member of its class. 
Pre-explosion {\it Hubble Space Telescope (HST)} imaging of the location of 
the SN, along with a precise localization of this event using the 
Laser-Guide-Star assisted Adaptive Optics (LGS-AO) system 
at Keck Observatory, are combined to identify a luminous ($M_V=-10.3$) point source
as the possible progenitor of SN 2005gl. If the source is indeed a single star, it was
likely a member of the class of luminous blue variable stars (LBVs). This 
finding leads us to consider the possible general association of SNe IIn with
LBV progenitors. We find this is indeed supported by observations of other SNe,   
and the known properties of LBV stars. For example, we argue that should the prototypical  
Galactic LBV $\eta$ Carina explode in a phase similar to its current state, 
it will likely produce a type IIn SN. 
We discuss our findings in the context of current
ideas about the evolution of massive stars, and review the census of SNe with identified 
progenitors. We introduce the concept of the progenitor-SN map as a convenient means to discuss the
present status and future prospects of direct searches for SN progenitors. We conclude that   
this field has matured considerably in recent years, and the transition from anecdotal
information about rare single events to robust associations of progenitor classes with
specific SN types has already begun.

\end{abstract}




\section{Introduction}

It is generally assumed that supernovae (SNe) can be divided into 
two physical classes. Type Ia SNe are assumed to result from the
thermonuclear explosion of a degenerate white dwarf star, reaching 
the critical ignition density as it approaches the
Chandrasekhar limit by accretion from, or merger
with, a binary companion. Direct observational evidence shows that
all other types of SNe result from the gravitational core-collapse
of young, massive stars. These progenitors are expected to be relatively luminous,
and are thus potentially detectable in images of sufficient 
spatial resolution and depth obtained before these core-collapse SNe explode.
 
The impact of the study of SN progenitors was poignantly illustrated 
by the watershed case of SN 1987A and its blue supergiant progenitor
(White \& Malin 1987). Initial surprise at the color (blue rather than red) 
and compactness of this progenitor led to revisions in our understanding 
of massive star evolution and SN explosion physics. During the next 15 
years progress was slow, with but a single additional progenitor 
identified (SN 1993J; Aldering, Humphreys, \& Richmond 1994; 
Van Dyk et al. 2002; Maund et al. 2004). 
Pioneering work by Van Dyk and collaborators (Barth et al. 1996; Van Dyk et al. 1999;
Van Dyk, Li, \& Filippenko 2003a) utilized a new 
resource - the sensitivity and resolution afforded by pre-explosion images
obtained by the Hubble Space Telescope ({\it HST}). Several possible 
progenitors have been identified, but these associations were often
inconclusive due to the SN astrometry being limited by 
post-explosion ground based images of relatively poor quality. 

In the last few years, breakthrough results were presented by two groups 
(the California group, e.g., Van Dyk et al. 2002; Van Dyk, Li, \& Filippenko 2003b, 
Li et al. 2006, and the UK group, e.g., Smartt et al. 2004; Hendry et al. 2006) using mostly 
post-explosion {\it HST} imaging to precisely determine the location of SNe 
and securely identify progenitor stars in pre-explosion {\it HST} images.

Most recently, we have introduced the use of laser guide star assisted adaptive
optics (LGS-AO) as an alternative means for precise SN localization that is 
independent of HST scheduling and operations, and not subject to saturation by the 
brightness of a young SN (Gal-Yam et al. 2005a,b). 
Here, we report the second result from our program at Keck
Observatory, the discovery of a luminous point source in pre-explosion images of 
the type IIn SN 2005gl. The paper layout is as follows. In $\S~2$ we present our
observations, including analysis of archival pre-explosion {\it HST} images of the location of this event,
and our Keck LGS-AO post-explosion observations leading to precise localization of
the SN on the pre-explosion grid. In $\S~3$ we present the discovery of a
point source consistent with being a very luminous LBV-type progenitor 
of this SN. We conclude in $\S~4$ with a discussion of our result in the 
context of accumulated information about SNe IIn and 
all available SN progenitor identifications to date. 

\section{Observations}

\subsection{Discovery and photometry}

SN 2005gl was discovered on 2005 Oct. 5.18 UT by Puckett \& Ceravolo, and independently by
Sano (Puckett et al. 2005). We identified this event as a type IIn SN (Blanc et al. 2005; $\S~2.4$) 
using a spectrum obtained with the SNIFS spectrograph mounted on the UH 2.2m
telescope (Fig.~\ref{specfig}) on 2005 Oct. 13.5 UT. Pre-explosion imaging 
(Puckett et al. 2005) places the explosion date of this SN between 2005 Sep. 10 and 2005 Oct. 5.    

\begin{figure*}
\includegraphics[width=15cm]{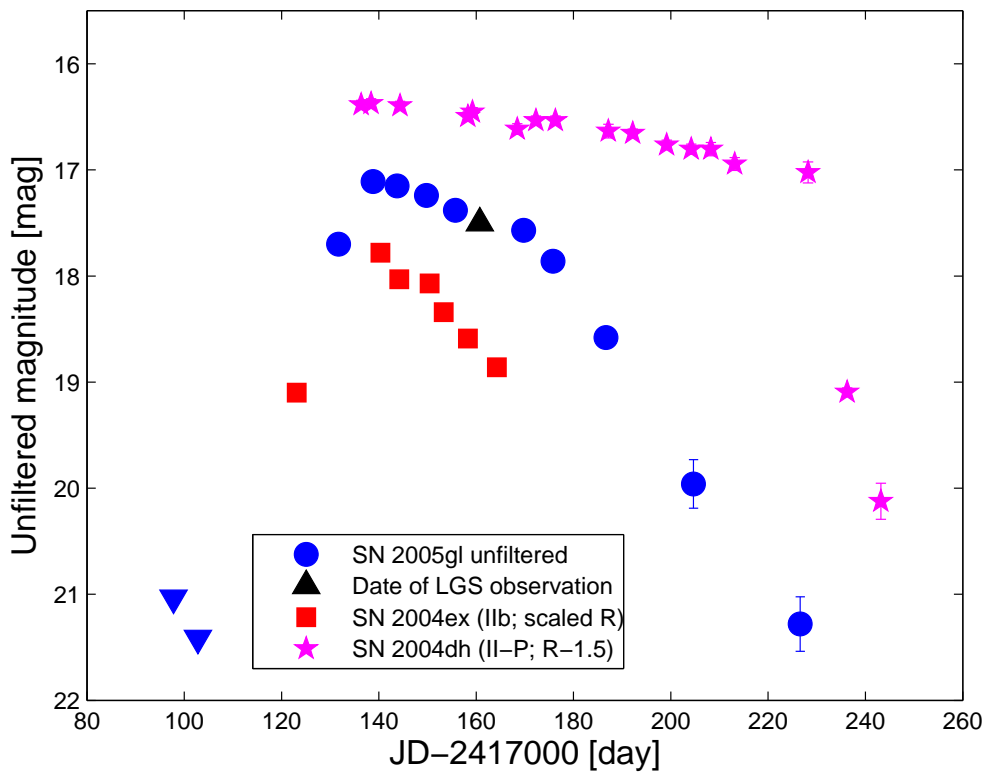}
\caption{A comparison of the scaled unfiltered KAIT light curve
of SN 2005gl (blue circles) with those of two SNe studied by the 
CCCP (Gal-Yam et al. 2004b; Gal-Yam et al. in preparation). 
SN 2004dh is a typical SN II-P, displaying the characteristic
long plateau in its R-band light curve (magenta stars; Nugent
et al. 2006; Gal-Yam et al. in preparation), while the R-band light 
curve of SN 2004ex (red squares; Gal-Yam et al. in preparation) shows
a rapid linear decline, typical for SNe IIb and II-L events. 
The light curve of SN 2005gl is intermediate, as is often seen for
SNe IIn. A possible interpretation is that the light curve 
represents the combination of a rapidly declining ``linear'' 
component, similar to those observed in other partially-stripped
SNe of types IIb and II-L, augmented by an additional
contribution from long-lasting circumstellar shocks, which 
cause the light curve to decline more slowly. The light curves 
were aligned to have approximately the same peak date, and
were arbitrarily scaled for clarity. Error bars are typically 
smaller than the symbol size, and upper limits are denoted by
inverted triangles.}
\label{photfig}
\end{figure*}

Unfortunately, the photometric coverage of this event is quite poor, and a light curve
in any standard filter cannot be derived from the data currently available to us. 
However, we are able to extract the light curve of this object from unfiltered 
survey images of the host galaxy,
routinely obtained by the Katzman Automatic Imaging Telescope (KAIT;
Li et al. 2000; Filippenko et al. 2001; Filippenko 2005) at Lick observatory. 
We use the image-subtraction-based photometry methods of Gal-Yam et al. (2004a)
to remove the underlying host galaxy light and measure the luminosity of this event. 
The unfiltered light curve has been placed on an $R$-band-equivalent zeropoint,
anchored to four nearby stars for which we obtained photometric calibration 
with the robotic 60 inch telescope at Palomar Observatory on July 26, 2006 UT 
(see Appendix).  
We find that SN 2005gl peaked at $R\approx17$ mag around Oct. 20, 2006 UT.
Fig.~\ref{photfig} shows a comparison between our light curve and those
of SN 2004dh, a typical SN II-P (Nugent et al. 2006; Gal-Yam et al. in preparation)
and SN 2004ex, a linearly-declining SN IIb (Gal-Yam et al. in preparation). As
is often seen in SNe IIn, the light curve of SN 2005gl does not show a long plateau
phase similar to those of SNe II-P, but it does declines quite slowly for $\sim50$ days,
compared to other linearly-declining events (SNe II-L/IIb). This slow decline is often
attributed to emission contributed by ongoing circumstellar shocks.  

\subsection{Spectroscopy}

Shortly after its discovery, on 2005 Oct. 13.5 UT, 
we observed SN2005gl with the SuperNova Integral
Field Spectrograph (SNIFS), a high-throughput dual-channel
lenslet-based instrument optimized for automated observation of point
sources on a diffuse background (Aldering et al. 2002). 
The single 1000s exposure was obtained close to zenith
(secz=1.1), and covers a fully-filled $6'' \times 6''$ field-of-view in the
$3300-10000$~\AA\ extended optical domain with a moderate spectral
resolution of 2.5~\AA\ between $3300-5100$~\AA\ and 3.4~\AA\ between 
$5100-10000$~\AA\ .

The IFS dataset was calibrated using a dedicated procedure
including CCD-preprocessing, diffuse-light subtraction, 3D-dataset
reconstruction (the extraction procedure uses an optical model of the
instrument to locate each of the spectra, and extracts them using
optimal weighting, taking account of the flux overlap between adjacent
spectra), wavelength-calibration and spectro-spatial flat-fielding
using internal calibration frames, cosmic-ray removal and telluric
feature correction.

The flux solution was computed from a spectrum of the white dwarf GD71 acquired
for 600s on the same night at equivalent airmass, using the standard
Mauna-Kea extinction curve (Beland, Boulade, \& Davidge 1988). The
night was reasonably photometric according to 
CFHT/Skyprobe\footnote{\url http://www.cfht.hawaii.edu/Instruments/Elixir/skyprobe/home.html}, 
with a stable seeing of 1'' FWHM.

\begin{figure*}
\includegraphics[width=9cm]{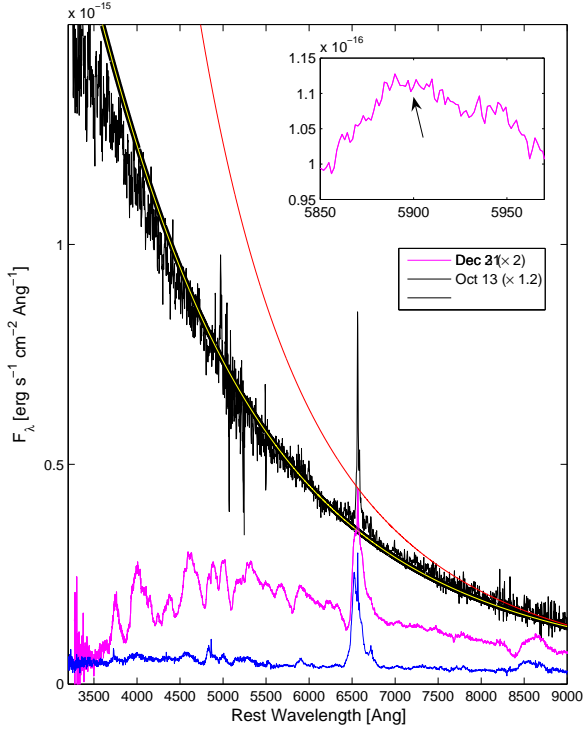}
\caption{Spectroscopy of SN 2005gl. The spectra show strong
Balmer H$\alpha$ and H$\beta$ lines, with a narrow core ($v\sim400$ km s$^{-1}$) 
superposed on a broad base ($v\sim2000$ km s$^{-1}$),
the hallmark of type IIn SNe. Our spectra do not
resemble those of the recently discovered hybrid Ia/IIn events such as SN 2002ic (Hamuy et al. 2003;
Wood-Vasey et al. 2004) and SN 2005gj (Prieto et al. 2005; Aldering et al. 2006). 
In particular, they show narrow Fe II lines (typical in spectra of SNe II) in the early Dec. spectrum,
and He I lines in emission (587, 667 and 706 nm) in the later one. 
They markedly lacks the prominent, broad Si and Fe line blends observed in SNe Ia and their hybrid IIn cousins
(see, e.g., Deng et al. 2004, Wang et al. 2004).
The inset shows a detailed view of the area around the wavelength of the Na I D line
in our highest S/N spectrum (Dec. 2). A
weak notch is detected at the correct wavelength (marked) but it is also consistent with
being a noise feature. We adopt the strength of this possible feature as an upper limit 
on the equivalent width of any real Na D absorption, and use this to constrain the
extinction toward this SN to be below $A_V=0.35$ mag. A robust (but less constraining, $A_V<1.05$) 
limit is obtained using black-body fitting to our earliest spectrum (yellow and red curves),
see $\S~3.2$ for details.}
\label{specfig}
\end{figure*}

The point-source spectra were extracted from the 3D-datasets using
an aperture-based algorithm (with an aperture radius of 4 and 5$\sigma$
for SN2005gl and GD71 respectively) after uniform background
subtraction (estimated outside the aperture). While the supernova lies on
a slightly structured host-galaxy background ($\sim 1\%$ RMS), we estimate the
overall accuracy of the SN spectrum flux solution to be
$\sim 5 \%$ everywhere but at the wavelength domain edges. Based on a preliminary
version of this spectrum the SN was initially identified as a type IIn event
(Fig.~\ref{specfig}; Blanc et al. 2005). 

We obtained further spectra of SN 2005gl on 2005 December 2.3 and 31.3 UT with the
Low-Resolution Imaging Spectrometer (Oke et al. 1995) double-spectrograph (in
polarimetry mode - LRISp - on Dec. 31) at the Cassegrain focus of the Keck-I
10-m telescope.  We used the dichroic filters D55(D68) to split the spectrum at
5500~\AA\ (6800~\AA) on Dec 2.3(31.3) respectively, and send the 
blue and red light to the respective arms of LRIS,
where the choice of gratings/grisms yielded a resolution on both sides of
$\sim 10$\AA\ and spectral coverage from about $3300 - 9100$\AA.  The total
exposure times were 1800s(600s) on Dec 2.3(31.3), and the object was observed at the
parallactic angle.

To derive the total-flux spectrum, we extracted the one-dimensional
sky-subtracted spectra optimally (Horne 1986) in the usual manner.  The
spectra were then wavelength and flux calibrated, corrected for continuum
atmospheric extinction and telluric absorption bands
(Wade \& Horne 1988; Bessell 1999; Matheson et al. 2000), and combined. Note that, although we
observed with LRISp on Dec. 31.3 UT, SN 2005gl was only observed with a single wave-plate
position, and thus polarization information is not available.

The spectral evolution of this object is presented in Fig.~\ref{specfig}. The earliest
SNIFS spectrum is very blue, and well-fitted by a black body (BB) spectrum with 
$T=13,000 {\rm K}$ (yellow curve). Assuing a very hot intrinsic (unreddened) spectrum 
(BB with $T=5 \times 10^5 {\rm K}$; red curve) we find that any dust reddening cannot
be stronger than $E_{B-V}=0.3 (A_V=1.05)$ without the resulting model spectrum (reddened BB) 
underpredicting the blue flux. $A_V=1.05$ therefore represents a robust upper limit
on the amount of possible dust extinction towards this object (see below).

By Dec. 2, the spectrum has evolved and is now dominated by Balmer and Ca lines with 
P-Cygni profiles and prominent Fe-II absorption lines near $5000$\AA\ typical to type
II SNe at this age ($\sim50$ days after explosion). The last spectrum (Dec. 31) is
dominated by emission lines of H and He, while absorption features are weak or have
disappeared.  

\subsection{Pre-explosion {\it Hubble Space Telescope} observations}

The host galaxy of SN 2005gl, NGC 266, was observed by {\it HST} in 1997 as part 
of a program to study nearby galaxies with active nuclei (GO 6837, PI Ho). 
Imaging was obtained in UV (F218W; data archive designation u3mj0101m 
and u3mj0102m, 900 seconds each) and V (F547M; u3mj0103m and u3mj0104m, 200 and 
160 seconds, respectively) bands using the Wide-Field
and Planetary Camera 2 (WFPC2) instrument. While the galaxy
nucleus (the target) was positioned on the PC chip, the location of SN 2005gl
was fortunately placed on wide-field chip 2 (WF2). Following the explosion
of SN 2005gl in October 2005, we have located and retrieved these data 
from the {\it HST} archive.

The frames were preprocessed through the standard Space Telescope Science
Institute pipeline using the latest calibrations as of 2005 Dec 24.  The images
were further processed using the suite of programs designed specifically for
the reduction of WFPC2 data that are available as part of the HSTphot
(Dolphin 2000) software package (version 1.1.5b; our implementation
includes all updates through 2003 May 28), following the procedure
outlined by Leonard et al. 2003. When possible, {\it hstphot} returns
magnitudes in standard Johnson-Cousins photometric bands as output.  
For our observations, the complete
transformation to standard {\it V} is not possible since color information is
not available (i.e., $I$-band observations were not taken); however, comparison
of the flight-system magnitudes with standard $V$ for a different dataset (see
below) reveals that the color-correction is generally under $0.03 {\rm\ mag}$
for most objects, so that our flight-system magnitude should be quite close to
the standard $V$ magnitude.  

\begin{figure*}
\includegraphics[width=15cm]{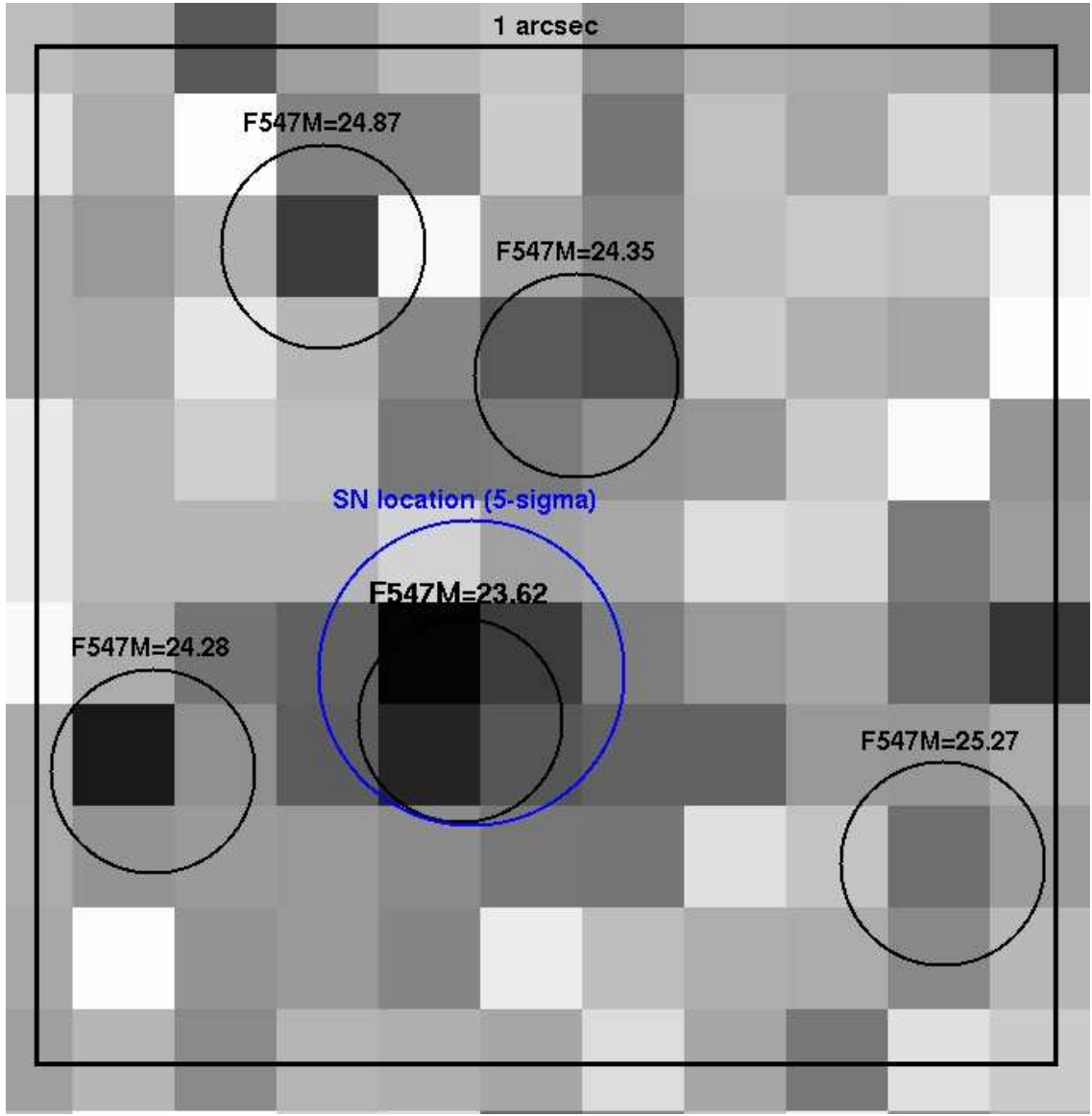}
\caption{Sources near the location of SN 2005gl as seen in
pre-explosion {\it HST} imaging. In black we mark the location and apparent 
magnitudes of the 5 point sources identified by HSTphot ($\S~2.1$ and Table 1). 
In blue we plot the 5$\sigma$ error circle around the location of the SN, as
determined from registration of high-resolution post-explosion Keck-AO images
onto the pre-explosion {\it HST} grid (see $\S~3.1$ and Fig.~\ref{LGShst}). A single
point source is consistent with being the progenitor of SN 2005gl. The {\it HST}
image section shown in approximately $1''$ on the side, north is up and east 
is to the left.}
\label{hstpre}
\end{figure*}

We ran {\it hstphot} with option flag 14, which combines turning on local sky
determination, turning off empirically determined aperture corrections (using
default values instead), and turning off PSF-residual determination; these are
the recommended settings for a galaxy well beyond the Local Group.  By turning
off empirical aperture corrections, default values for each filter are applied
to the photometry, and are accurate, in general, to 0.02 mag.  {\it Hstphot}
was run with a S/N threshold of 1.0.
We identify several point sources near the location of the SN on WF chip 1 
(Fig.~\ref{hstpre} and Table 1). 

Photometry produced by HSTphot has been compared with that generated by other
packages, including DoPHOT (Schechter, Mateo, \& Saha 1993) and DAOPHOT/ALLFRAME
(Stetson 1987, 1994), and the results show excellent agreement
(Dolphin 2000).  This is not surprising, since most of the
machinery of HSTphot, including the star-finding and PSF-fitting algorithms,
are in fact modeled after these older packages. The primary advantage of
HSTphot is that it has built-in knowledge of WFPC2 instrumental characteristics
and hence runs with far less user interaction and produces robust results that
are easily reproducible by different users.

However, as far as we are aware, no direct comparisons exist in the literature
between HSTphot and other photometry packages for WFPC2 observations using the
F547M filter.  As a ``sanity check'' on our implementation of HSTphot, then, we
reduced and analyzed the archival {\it HST} WFPC2 data acquired for the Small
Magellanic Cloud as part of a stellar populations study (program GO-8196) using
the F547M and F814W filters (datasets u5ct0201r, u5ct0202r, u5ct0203r,
u5ct020dr, u5ct020er, and u5ct020fr).  A comparison between our {\it hstphot}
$V$ and $I$ magnitudes with the values derived by McCumber et al. (2005) using
the IRAF task APPHOT.PHOT for a sample of stars listed in Table 8 of McCumber
et al. (2005) yields overall agreement to within 0.06 mag in both filters,
building confidence in the robustness of our results.

\begin{deluxetable}{llllll}
\tabletypesize{\scriptsize}
\tablecaption{Point sources near the location of SN 2005gl}
\tablewidth{17cm}
\tablehead{
\colhead{\#} & \colhead{Chip position (x,y)} & \colhead{Signal-to-noise ratio} & \colhead{Counts} & \colhead{Magnitude (flight system)} & \colhead{Error (mag)}\\
}

\startdata

1 & (56.72,157.95) &  7.2 & 36.401 & 24.036 & 0.150 \\ 
2 & (53.67,157.22) &  4.8 & 21.295 & 24.607 & 0.226 \\
3 & (58.03,161.12) &  4.3 & 19.427 & 24.702 & 0.255 \\
4 & (55.47,162.42) &  2.4 & 12.922 & 25.133 & 0.455 \\
5 & (61.63,156.34) &  3.0 & 12.198 & 25.195 & 0.357 \\

\enddata

\tablecomments{The X and Y positions reported by
HSTphot follow the convention that an integer value is assigned to a star that is
centered in the lower left corner of a pixel; this is similar to the output from
DoPHOT, but 0.5 lower in both X and Y than DAOPHOT.}
\end{deluxetable}

\subsection{Keck Laser-Guide-Star assisted Adaptive Optics Observations}

We observed SN\,2005gl on the night of 11~November 2005 UT with the
wide-field channel (plate scale 0.04\arcsec\ pixel$^{-1}$) of the Near
Infra-Red Camera 2 (NIRC2) operated behind the Laser Guide
Star-Assisted Adaptive Optics System (LGSAO; Wizinowich et al. 2006)
on the Keck~II \mbox{10-m} telescope on Mauna Kea, Hawaii.  
At that time ($\sim20$ days after peak, Fig.~\ref{photfig})
we estimate the brightness of the SN was $K\approx 16 \pm 0.2$\,mag in
the NIR (see Appendix for calibration details)
and $R\approx 17.5$\,mag in the optical (Fig.~\ref{photfig}). We observed for
25 $\times$ 30\,s in a 5-point box dither pattern through the $K_p$
(2.1 $\mu$m) filter; the mean epoch of our observations was 07:40 UT.

Calibration products included bias images, afternoon dome flats taken
both with and without dome-lamp illumination -- to enable subtraction
of the underlying thermal signature -- and a bad pixel map, initially
derived from the flatfields and then refined during subsequent
analysis.  After bias-subtraction and flat-fielding, the sky
background of individual science frames was estimated and used as a
normalization for the purpose of calculating a single fringe image.
Note that the domination of the sky background at NIR wavelengths by
bright emission lines inevitably produces fringes, and that since
fringes are an additive background it is not strictly appropriate to
derive flat fields from night sky images in the NIR.

Fringe subtraction was followed by image registration (shift and add),
cosmic ray identification, refinement of the bad pixel mask, cosmic
ray and bad-pixel masking, and image combination (imcombine) managed
using custom software within the Pyraf environment.  The resulting
coadded image has a full-width at half maximum (FWHM) of 0.10\arcsec\
(2.6 pixels), with a peak signal of 8000~DN from SN\,2005gl, which is
well below detector and ADC (32-bit) saturation.  The overall cosmetic
quality of the image is not ideal, mainly because diffuse emission
from the host galaxy NGC\,266 was incorporated into the fringe image,
causing misestimation of the sky background in individual frames, with
this effect exacerbated by the simple dither pattern.  However, these
defects are not expected to impact the astrometric utility of the
image, and were judged to be sufficiently negligible within the region
of interest that a more refined analysis of the data was not
attempted.

\begin{figure*}
\caption{Registration of the post-explosion $K_p$-band Keck-LGS image (top panel)
onto the pre-explosion $V$-band {\it HST} image (bottom panel). 8 common nearby compact
sources are identified and outlined. The 3 sources marked in magenta were excluded
from the final fits (see text). The SN location is determined from the unsaturated 
Keck image using centroiding algorithms within IRAF, and projected onto the pre-explosion
{\it HST} grid using the geometric solution. The red circle marks the 5$\sigma$ 
error circle with the uncertainty accounting for both the estimated centroiding 
and registration errors. (Removed due to arXiv limits. See JPEG version of this figure)}
\label{LGShst}
\end{figure*}

\section{Results}

\subsection{A possible luminous progenitor for SN 2005gl}

We have registered the post-explosion Keck-LGS images to pre-explosion
{\it HST} images following the procedures described in Gal-Yam 
et al. (2005a). Briefly, the process included the identification of nearby compact
sources (marked in Fig.~\ref{LGShst}), detected in both images, which are used to calculate
the geometric registration solution using the task {\it geomap} within IRAF. 
The final solution we obtained had an RMS residual of $\sim0.33$ pixel in X
and $0.36$ pixel in Y, using the 5 stars circled in black in Fig.~\ref{LGShst}. In    
magenta we mark 3 additional sources that match well 
but were excluded from the final fit due to being too faint in one 
of the images, or elongated with mismatched centers. 
Registration using various subsets of the black and magenta 
sources yield consistent results. 
The small overlap area between the {\it HST} and Keck-LGS images, 
and small number of available common sources, does not require or justify
high order geometric solutions, so we have solved only for a shift and 
scale correction between the distortion-corrected {\it HST} and Keck-LGS frames.

The blue circle in Fig.~\ref{hstpre} shows the 5-sigma error circle (with radius $\sim 0.06''$) 
around the localization of SN 2005gl,
as derived from the HST-LGS registration demonstrated in Fig.~\ref{LGShst}. The SN location was
determined using the centroiding algorithm within IRAF. The centroiding error was estimated
by comparing centroid positions obtained using different algorithms and extraction apertures
around the SN location, and found to be negligible compared to the uncertainty introduced
by the geometric solution. The final uncertainty reported above and presented in Fig.~\ref{LGShst} 
was calculated by adding the centroiding and registration uncertainty in quadrature. 

A single point source is consistent with the SN location. 
We measure a flight-system magnitude of $24.04 \pm 0.15$ for this object,
(statistical error only). Although we are unable to make a formal 
color-correction to translate this flight-system magnitude into standard $V$, 
we believe this value is within 0.03 mag of the $V$ magnitude ($\S~2.1$).  
Adding a conservative value of 0.05 mag to the Poisson uncertainty yields a final, 
$V$-band magnitude of $24.04 \pm 0.16$ mag.

Since the host is relatively distant ($z=0.015547 \pm 0.000017$, Huchra, Vogeley, \&
Geller 1999, via NED\footnote{\url http://nedwww.ipac.caltech.edu/}) we use Hubble's law to calculate a distance of 
$66 \pm 4$ Mpc (for H=70 km s$^{-1}$ Mpc$^{-1}$) to this galaxy, where the error is dominated 
by the effects of our adopted peculiar velocity uncertainty,  
$v=\pm300$ km s$^{-1}$. The distance modulus is $\mu=34.1 \pm 0.15$ and thus
the absolute magnitude of the putative progenitor is $M_V=-10.3 \pm 0.2$, where we
also account for $A_V=0.23$ magnitudes of galactic extinction, as derived from
the Schlegel, Finkbeiner \& Davis (1998) dust maps. If this is a single
star, it most likely belongs to the class of luminous blue variables
(LBVs, with bolometric absolute magnitudes above $M=-9.5$), to which belong
all known stars of such high luminosities, including all the well-studied cases
in our Galaxy (see below). Red supergiants (RSGs; $M>-9$), blue supergiants (BSGs; e.g., the 
progenitor of SN 1987A, $M\sim-8$) and even the so-called rare ``cool hypergiants''
($M>-10$) are all fainter than this source (see, e.g., Humphreys \& Davidson 1994;
their Fig. 9). 
    
\subsection{Caveats and future prospects}

The magnitude given above is not corrected for possible extinction in NGC 266, the host
galaxy of SN 2005gl. Since objects near the SN location are detected only in 
a single band in the pre-explosion image, they do not constrain the amount 
of extinction close to this line of sight. However, the spectroscopy of the SN
itself indicates this value is small. Our data do not show strong Na D 
absorption lines (Fig.~\ref{specfig}, inset), which are correlated with dust extinction (see Turatto et al. 2003
for the latest compilation). The notch seen at the location of the Na D line in our spectrum
is consistent with a noise feature. Adopting its strength as an upper limit
on any real Na D absorption, the measured equivalent width (EW$=0.03$ \AA) implies
$E_{B-V}=0$ using the latest formulas from Turatto et al. (2003). As can be seen
in Fig. 3 of that work, all SNe with Na D lines weaker than EW$=0.2$ \AA~ have
measured $E_{B-V}\lesssim0.1$, which we therefore adopt as a conservative upper limit 
on dust extinction along this line of sight, implying $A_V<0.35$ mag. 

Additionally, the very blue spectrum of the SN at early time
argues against significant extinction. We follow Leonard et al. (2002)
in quantifying this observation in the following manner. Fig.~\ref{specfig} shows that 
the early spectrum of SN 2005gl is well-fit by a black body spectrum with
$T=13000 {\rm K}$ (yellow curve). Also plotted is the black body curve
for a source with $T=500000 {\rm K}$, representing the hottest intrinsic
spectrum one can suggest for a SN a few days after explosion (red curve). 
Reddening this putative hot spectrum using the Cardelli, Clayton \& Mathis
(1989) law with R=3.08 and extinction values of $E_{B-V}>0.3$ results in model observed
spectra that are redder (underpredict the blue flux) than our SNIFS spectrum.
We therefore conclude that $E_{B-V}=0.3$ ($A_V\approx1.05$) is a robust upper
limit on the optical extinction of SN 2005gl (the likely value, as argued above,
is much smaller).   
Note, that such modest host extinction would drive up the magnitude of 
the putative progenitor star, making an LBV identification stronger yet. 
 
The main caveat remaining is establishing that the point source we have detected in 
the pre-explosion {\it HST} imaging is the progenitor of SN 2005gl is the 
possibility that this source is a compact, luminous star cluster, rather than
a single star. Compact luminous clusters with properties similar to those
we measure (often called ``super star clusters'', or SSCs) are observed in very
active galaxies (such as starburst or interacting galaxies). However, they are generally
rare in more normal galaxies (with but a single SSC candidate, Westerlund 1, observed in
our galaxy) and are expected to be even less frequent in earlier-type galaxies such as
NGC 266 (an Sa galaxy, Nilson 1973; via NED). The majority of massive SN progenitors in early spirals are not 
found in SSCs, and thus we believe that SN 2005gl probably did not explode in such an
environment. However, this possibility cannot be ruled out at this time. Additional 
{\it HST} imaging obtained once SN 2005gl declines should provide decisive evidence,
with the point source at the SN location either gone (if it was a single star) or
remaining approximately at the same magnitude, if it is indeed an SSC. 

Additional support for the association between SNe IIn (such as SN 2005gl) and LBV progenitors
could arise if additional such cases are discovered. While SNe IIn are intrinsically
rare (Cappellaro et al. 1997), they tend to be over-represented in observed SN sampled due
to their high average luminosity. In addition, the luminosity of LBV progenitors
makes them visible in galaxies which are far more distant than hosts of less 
luminous stars. The number of such galaxies in the {\it HST} archive is higher
than the number of nearby ones ($d<20$ Mpc) usually considered as likely candidates for progenitor studies,   
again increasing the opportunity to test the association of SNe IIn with
LBV progenitors in the future. 

\section{Discussion and Conclusions}

\subsection{SNe IIn from LBVs}

We now examine the hypothesis that LBV stars explode as SN IIn. 
The class of type IIn SNe is known to be a heterogeneous group of events. 
At least two subsets represent specific and probably unrelated phenomena to the one
we consider here, namely, the core-collapse-driven explosion of a massive
LBV star, similar, for example, to the well-known $\eta$-Carina in our
Galaxy. The first unrelated class are the so-called ``SN impostors'', which
are believed to be super-outbursts of LBVs which do not result in total 
disruption of the progenitor star (see, e.g., Van Dyk 2005 and Maund et al. 2006, and references
therein). These events are typically 
faint (compared to a normal SN) with absolute magnitudes between 
$-10$ and $-14$ (Van Dyk 2005), while ``genuine'' type IIn explosions (Schlegel 1990), 
are much brighter (sometime reaching $M=-20$; SN 2005gl has $M\sim-17$). 
The second group of events which are also 
probably irrelevant to our discussion are the recently discovered class
of ``hybrid'' Ia/IIn SNe, including the prototype SN 2002ic 
(Hamuy et al. 2003; Wood-Vasey et al. 2004) along
with its recent (SN 2005gj, Prieto et al. 2005; Aldering et al. 2006) and past (SN 1997cy, Germany et al. 2000, Turrato et al.
2000; SN 1999E, Rigon et al. 2003) clones (Hamuy et al. 2003, Deng et al. 2004, Wang et al. 2004).
These SNe display spectral properties similar to those of thermonuclear SNe Ia,
along with narrow Balmer lines that probably result from strong interaction with
circumstellar material in the immediate vicinity of the exploding star. 
Such events have been suggested to be thermonuclear SNe Ia, exploding either
in close proximity to recently stripped gas from a binary companion with
intense mass loss (e.g., an AGB star, Hamuy et al. 2003); within a
symbiotic system; or perhaps thermonuclear explosion that occur while the
envelope of the exploding star is still intact (''SNe 1.5'', Iben \& Renzini 1983). 
In any case, SN 2005gl does not share the spectroscopic properties of these hybrid
events, and does not appear to be related to this IIn sub-class (Fig.~\ref{specfig}). We note that
both of the sub-groups discussed above comprise a minority within the observed
population of SNe IIn.  
     
So, we can reformulate our question to be: can LBVs be the
progenitors of most SN IIn explosions? We focus on the best-studied example of an LBV
in our Galaxy, $\eta$ Carina. The envelope of this star still contains large quantities 
of hydrogen (e.g., Davidson et al. 1986), so, had it exploded now, it would result in a type
II (H-rich) SN. In addition, we observe the results of copious mass loss around this
system, of order several solar masses of ejected material, mostly H. Were an energetic explosion
to occur at the center of such a huge debris cloud, we would expect strong interaction leading to
strong narrow H lines - the hallmark of type IIn SNe. Thus, we conclude that
an LBV exploding during the active mass-ejection phase, or shortly thereafter,
would indeed appear to distant observers as a type IIn SN.

The rates of SNe~IIn may shed some light on our proposed progenitor association.
Let us consider a simplistic massive star evolutionary scheme,
for single stars with approximately solar metallicity, broadly following, e.g., Maeder and Conti (1994).
We use the customary notations RSG for red supergiants, WN for N-rich (and usually also He-rich) Wolf-Rayet 
stars, WC for C-rich W-R stars, and WO for O-rich W-R stars. 

\begin{equation}
80 M_{\odot}< M < 150 M_{\odot}: O \rightarrow LBV \rightarrow {\rm SN IIn} (?)
\end{equation}
\begin{equation}
40 M_{\odot}< M < 80 M_{\odot}: O \rightarrow LBV \rightarrow WN \rightarrow WC/WO \rightarrow {\rm SN Ic}
\end{equation}
\begin{equation}
25 M_{\odot}< M < 40 M_{\odot}: O \rightarrow LBV \rightarrow (early) WN \rightarrow {\rm SN Ib}
\end{equation}
\begin{equation}
15 M_{\odot}< M < 25 M_{\odot}: O \rightarrow RSG \rightarrow (late) WN \rightarrow {\rm SN IIL/IIb}
\end{equation}
\begin{equation}
8 M_{\odot}< M < 15 M_{\odot}: B/O \rightarrow RSG \rightarrow {\rm SN IIP}
\end{equation}

We note that SNe IIn can, in this picture, occur in two distinct cases. First, it is believed that
massive stars undergo short LBV phases, involving rapid and strong mass loss 
which transform H-rich supergiants to W-R
stars, as well as during the early W-R evolution stages (while the stars still have significant 
amounts of H, i.e., are WN stars; see, e.g., Maeder \& Conti 1994). 
Furthermore, since the evolution of the inner core is decoupled from that of the 
envelope during these stages, in some cases the core might collapse 
during those short-lived transition phase, resulting in SNe IIn from stars in the mass range,
comparable to that of the progenitors of SNe IIb/L and Ib/c. 
In this context we note the recent work by Chugai \& Chevalier
(2006) interpreting the observations of SN 2001em, which exploded as a SN Ic, and then
developed strong interaction signatures (a bright radio signal accompanied by SN IIn-like 
narrow H$\alpha$ lines, Stockdale et al. 2004; Soderberg, Gal-Yam, \& Kulkarni 2004; 
Bietenholz \& Bartel 2005). Models by these authors suggest that SN 2001em underwent
a violent and intense, hydrogen-rich, mass ejection episode, shortly ($\simlt 1000$) years before
the SN explosion. It appears SN 2001em exploded shortly after it exited a recent LBV phase, and
a similar event, occurring slightly (on stellar evolution timescales) earlier, would
appear as a SN IIn.  

The second case is a speculative option which we now describe. We hypothesize that in the case
of the most massive stars (Eq. 1) the evolution of the core might be rapid enough to overtake the progress 
of envelope mass loss, and that these stars might undergo core-collapse while in their first LBV
phase, before they are significantly stripped. SNe IIn in this case take the place of SNe Ic
as the end products of the rare, most massive stars. It is interesting to note in this 
context that explosion models of SNe Ic, including the most energetic ones associated with
GRBs (e.g., Deng et al. 2005; Mazzali et al. 2006) 
require massive progenitors (with zero-age mass $\simlt50 M_{\odot}$), 
but not as massive as the claimed masses of the most extreme LBVs ($>80 M_{\odot}$), e.g.,
$\eta$-Carina (Figer et al. 1998), LBV 1806 (Eikenberry et al. 2004) or the Pistol star (Figer et al. 1998).
We note that this scenario, in which the core evolution overtakes the envelope 
stripping, is not in accord with current stellar evolution models. Still, we
think it is an interesting option that merits further theoretical and observational 
examination. 

We note that Chugai et al. (2004) interpret their observations of the type IIn SN 1994W
as indicating that the progenitor underwent an explosive mass ejection shortly before its 
core collapsed and led to its final explosion as a SN. This would fit well with the above picture
of SNe IIn resulting from LBVs like $\eta$-Carina (which are known to undergo extreme and violent massive loss
episodes) during or shortly after their active phase.  
   
In either of the above cases, we expect these SNe from LBV progenitors 
to be rare (either because LBVs are short-lived phases, or 
because the progenitors are very rare stars). For example, the total number of stars with initial mass
above $80 M_{\odot}$ is $\sim3 \% $ of the total core-collapse progenitor population
($8M_{\odot}<M<150M_{\odot}$) for a Salpeter (1955) initial mass function.  
This is similar to the estimated fraction of SNe IIn of the total core-collapse
population ($\sim2 \%$, Cappellaro et al. 1997). 
We conclude that the notion that some or most of SNe IIn result from explosions of 
LBVs appears broadly consistent with available observational data.  

\subsection{SN 2005gl in context: The SN-Progenitor map}

\begin{figure*}
\includegraphics[width=15cm]{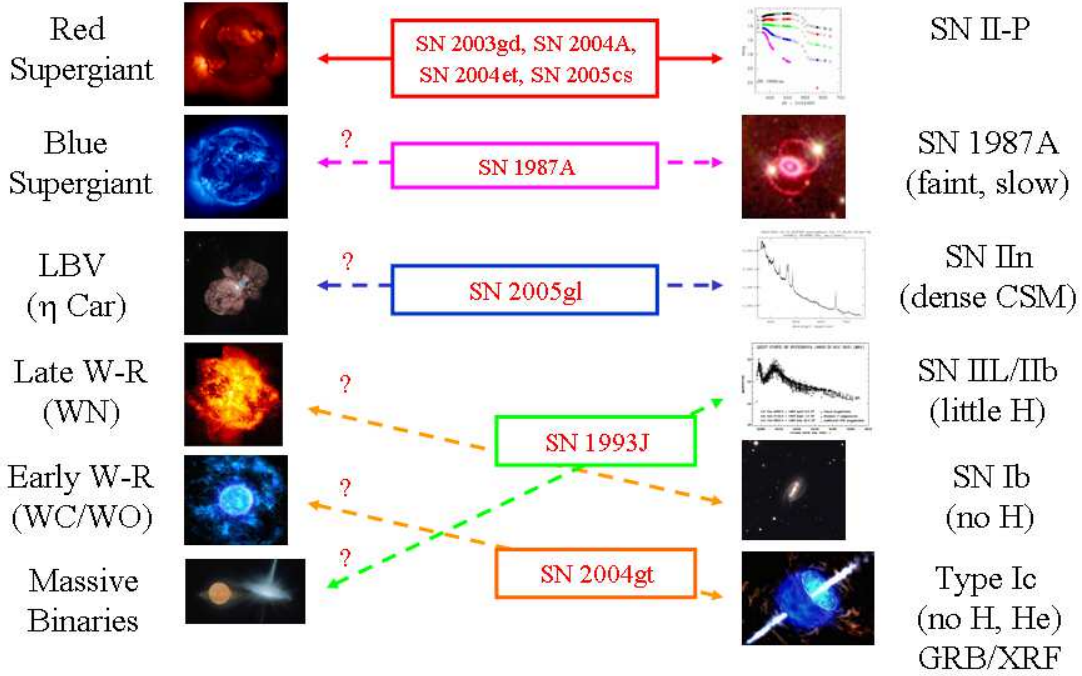}
\caption{The progenitor-SN map, presenting associations of SNe with progenitor stars 
based on direct observations of the progenitors in pre-explosion images. 
The association of the most common type of core-collapse events, SNe II-P,
with (relatively) low-mass ($8 M_{\odot}< M <15 M_{\odot}$) red supergiants (RSGs)
appears to be quite robust. The connections between fainter SN II events (1987A-like)
and the type IIn sub-class, with blue supergiants and LBVs, respectively, is but suggestive,
based on a single event in each case. The progenitors of stripped-envelope events (type Ib/c) 
are yet to be observed, but analysis of available upper limits suggest that the progenitors
of the more highly stripped (type Ic) SNe are more evolved (and less luminous) ``early'' W-R stars,
while SNe Ib, which still retain much of their He envelope, result from somewhat less evolved
(and often more luminous) ``late'' W-R stars.   
Data used are from White \& Malin 1987; Aldering et al. 1994; Van Dyk et al. 2002; Maund et al. 2004; 
Smartt et al. 2004; Van Dyk et al. 2003b; Li et al. 2005; Maund et al. 2005a; Li et al. 2006; 
Hendry et al. 2006; Gal-Yam et al. 2005a; Maund et al. 2005b; and this work.
}
\label{mapfig}
\end{figure*}

In Fig.~\ref{mapfig} we summarize the accumulated knowledge from direct observations of SN progenitor stars 
in the form of a progenitor-SN map. As possible progenitor classes we consider red and blue
supergiants (RSGs and BSGs), LBVs, early (He rich) and late (C/O rich) Wolf-Rayet
(W-R) stars, and massive binaries. On the SN side we list all the major well-defined
sub-classes of core-collapse SNe. Among the H-rich type II subclasses, SNe II-P have extended
plateau optical light curves, while SN 1987A-like events are fainter and have a late 
hump-like peak in their optical light curves. SNe IIn (excluding the specific subsets
described in $\S~4.1$ above) generally have blue and rather featureless spectra with Balmer emission-line
profiles which include a narrow component ($<1000$ km s$^{-1}$) and are often very
luminous in the optical, consistent with a strong contribution from shocked CSM. The
transition class of SNe IIL/IIb have rapidly declining light curves and often display 
strong He lines, developing into events with spectra similar to SNe Ib (see below). Of the 
H-poor type I SNe, type Ib events are dominated by He lines, while type Ic lack both
H and He and are dominated by intermediate-mass elements such as O, Mg, Si and Ca. 
It must be stressed that this is, by necessity, a simplified presentation of the core-collapse
SN ``zoo'', and that many peculiar objects which do not strictly fit into this 
picture, or transition objects that are either intermediate or evolve between 
classes, are known. Still, this scheme should include the majority of observed objects.     

We suggest this is a useful tool to present and discuss 
this topic. Of course, we have no guaranties that this map is complete (perhaps more 
progenitor classes or SN types are required) nor that it is a 1-to-1 mapping. Most likely this
version will evolve and change. Let us use it for the time being as a basis for 
discussion. We note that the association of type II-P SNe with low-mass 
RSG progenitors is well-supported by 4 observed events, with no counter examples,
and maintain this association is now progressing from suggestive to robust
(including also numerous supporting upper limits; Smartt et al. 2006).   
Studies of the (H and He-poor) type Ic SN 2004gt (Gal-Yam et al. 2005; Maund et al. 2005) 
and to a lesser degree SN 2002ap (Smartt et al. 2002) suggest these had highly evolved 
(i.e., optically less-luminous) W-R progenitors, perhaps of the WC or WO subclasses. 
This may fit well with the population of less-evolved W-R stars (late WN) 
exploding as He-rich type Ib events, consistent with recent explosion models by 
Tominaga et al. (2005). This picture is consistent with that expected from stellar 
evolution theory. The case of the transition class of type IIb objects (which evolve from having H-dominated
type II spectra toward the He-dominated type Ib class) is less clear. Observations of the
prototypical event SN 1993J in M81 suggest a possible massive binary progenitor (Aldering et al. 1994;
Van Dyk et al. 2002; Smartt et al. 2004) with some support from the recent work on 
SN 2001ig (Ryder et al. 2004; 2006). However, it is unclear whether the nearby putative 
binaries were indeed physically important or even bound to the SN progenitors,
or perhaps (some) SNe IIb arise from single WN (H rich) W-R stars (Soderberg et al. 2005).
 
Finally, we note that type IIn events, which show evidence for intense interaction with
circumstellar material, suggesting heavy mass-loss from the progenitor,
appear to require a different progenitor from those so far
discussed. Red supergiants have little mass loss, and so had the blue supergiant
progenitor of SN 1987A. While LBVs are not the only option, they provide an elegant solution, as discussed
above. It would appear that our putative map now encompasses all main SN types. Of course, more data 
are required to establish a robust mapping, and we would like each line to be supported by several progenitor-SN 
associations (as is the case for SNe II-P and red supergiants). Additional
cases may also allow an attempt to understand the effects of other parameters (beyond initial mass) such
as metallicity, rotation and binarity. 

\subsection{Conclusions}

We have identified a luminous point source in pre-explosion images of the location of the typical
type IIn SN 2005gl. If this was a single star it was most likely an LBV, based on luminosity considerations. 
We have discussed a major caveat - the
possibility that this point-like source is actually a compact luminous cluster - and 
ways by which this progenitor identification can be tested
by future observations. We suggest that an association of SNe IIn with LBV progenitors 
accommodates much of what is currently known about both LBVs and SNe IIn.
We introduce the progenitor-SN map and use it as a basis to discuss our
finding. This field has rapidly progressed in recent years and should evolve from a collection of
isolated fragmented pieces of information to become based on 
more solid concepts (such as the progenitor-SN map). These should soon begin to provide 
useful constraints on models, and drive additional progress in understanding the last stages
of massive star evolution and the physics of SNe.

\section*{Acknowledgments}

We are grateful to V. Dwarkadas, R. de Grijs, D. Maoz, P. Nugent, E. O. Ofek, S. Smartt 
and N. Smith for useful advice. K. Dawson and K. Barbary are thanked for assistance with 
LRIS observations at Keck.
This research has made use of the NASA/IPAC Extragalactic Database (NED) 
which is operated by the Jet Propulsion Laboratory, California Institute of Technology, 
under contract with the National Aeronautics and Space Administration.
A.G. acknowledges support
by NASA through Hubble Fellowship grant \#HST-HF-01158.01-A awarded by
STScI, which is operated by AURA, Inc., for NASA, under contract NAS
5-26555. A.G. further acknowledges the hospitality of the community of Cefalu and the efforts of the
organizers of the 2006 Cefalu international astronomy conference, during 
which this work has come to fruition. 
D.C.L. acknowledges support from an NSF Astronomy and Astrophysics
Postdoctoral Fellowship (award AST-0401479), under which part of this work
was completed.
D.J.S. acknowledges support provided by NASA through Chandra Postdoctoral
Fellowship grant number PF5-60041.
The work of A.V.F.'s group at UC Berkeley is supported by National Science
Foundation (NSF) grant AST-0307894, as well as by NASA grant AR-10690 from the
Space Telescope Science Institute, which is operated by AURA, Inc., under NASA
contract NAS5-26555. KAIT and its ongoing research were made possible by
generous donations from Sun Microsystems, Inc., the Hewlett-Packard Company,
AutoScope Corporation, Lick Observatory, NSF, the University of California, the
Sylvia \& Jim Katzman Foundation, and the TABASGO Foundation.
This work was supported in part
by the Director, Office of Science, Office of High Energy and Nuclear
Physics, of the U.S. Department of Energy under Contracts No.
DE-FG02-92ER40704, with additional support from the Gordon \&
Betty Moore Foundation and the CNRS/IN2P3, CNRS/INSU and
PNC agencies in France.

\newpage

\begin{center}
    {\bf Appendix: photometric calibrations}
\end{center}

We calibrated the field of SN 2005gl in $VR$ using observations
of Landolt (1992) standard fields (PG 1657+078, PG 0231+051 and SA 95)
obtained with the robotic 60 inch telescope at Palomar Observatory
on the night of July 25, 2006. The IRAF procedure {\it meastan} by
D. Maoz was used to calculate photometric solutions, which are
well-behaved, indicating the night was probably photometric. 
The overall zeropoint calibration error is $\sim 5\%$.
Table~\ref{phottable} and Fig.~\ref{chartfig} report the locations and 
magnitudes of stars near the SN location. 

The following equations were used to calculate the calibrated magnitudes:
\begin{equation}
R=-2.5 \times log(counts/s) + 22.112 - airmass \times 0.136 - (V-R) \times 0.065
\end{equation} 
\begin{equation}
V=-2.5 \times log(counts/s) + 21.952 - airmass \times 0.222 + (V-R) \times 0.023
\end{equation}

We calibrated the $K_p$-band magnitude of SN 2005gl in our NIRC2 LGS image 
by bootstrap calibration, based on two
common objects detected in the deep and narrow NIRC2 image and 
in a wider, more shallow $K_s$-band image
of the area obtained with the Wide-field Infra Red Camera (WIRC) mounted on the
200 inch Hale telescope at Palomar Observatory on July 20, 2006 (UT).
The WIRC image was reduced using the CCCP IR pipeline, in a similar manner 
to that described above for the NIRC2 data, and calibrated against the 2MASS catalog using
the methods described, e.g., in Gal-Yam et al. (2005a). 

\begin{figure*}
\caption{Field calibration. Nine stars close
to the host galaxy of SN 2005gl, NGC 266, with
magnitudes list in Table~\ref{phottable} are circled.
Only stars 2,5,6 and 7 appeared in the smaller KAIT
field of view and were used to derive the light 
curve of this object. (Removed due to arXiv limits. See JPEG version of this figure)
}
\label{chartfig}
\end{figure*}

\begin{deluxetable}{lllll}
\tablecaption{Calibration stars near SN 2005gl}
\tablewidth{17cm}
\tablehead{
\colhead{Star \#} & \colhead{R.A. (2000)} & \colhead{Declination} & \colhead{V [mag]} & \colhead{R [mag]}\\
}

\startdata

1 & 00:49:57.24 & +32:18:53.1 & 15.57 & 15.36  \\ 
2 & 00:49:52.45 & +32:18:43.4 & 16.51 & 16.39  \\ 
3 & 00:49:43.61 & +32:19:01.0 & 14.40 & 14.26  \\ 
4 & 00:49:43.74 & +32:18:48.0 & 14.76 & 15.02  \\ 
5 & 00:49:44.32 & +32:18:10.0 & 15.43 & 15.23  \\ 
6 & 00:49:36.00 & +32:18:01.6 & 17.02 & 17.06  \\ 
7 & 00:49:40.69 & +32:14:20.7 & 16.53 & 16.75  \\ 
8 & 00:49:59.84 & +32:14:23.0 & 15.81 & 15.64  \\ 
9 & 00:49:59.21 & +32:16:57.4 & 17.56 & 18.44  \\

\enddata

\label{phottable}
\end{deluxetable}


\begin{thebibliography}{}

\bibitem{} Aldering, G., Humphreys, R.~M., \& Richmond, M.\ 1994, \aj, 107, 662 
\bibitem{} Aldering, G., et al.\ 2002, \procspie, 4836, 61 
\bibitem{} The Nearby Supernova Factory Collaboration: G.~Aldering, et al.\ 2006, ApJ, in press, ArXiv Astrophysics e-prints, 
arXiv:astro-ph/0606499 
\bibitem{} Barth, A.~J., van Dyk, S.~D., Filippenko, A.~V., Leibundgut, B., \& Richmond, M.~W.\ 1996, \aj, 
111, 2047 
\bibitem{} B{\`e}land, S., Boulade, O., \& Davidge, T.\ 1988, Bulletin d'information du telescope Canada-France-Hawaii, 19, 16 
\bibitem{} Bessell, M.~S.\ 1999, \pasp, 111, 1426 
\bibitem{} Bietenholz, M.~F., \& Bartel, N.\ 2005, \apjl, 625, L99 
\bibitem{} Blanc, N., et al.\ 2005, The Astronomer's Telegram, 630, 1 
\bibitem{} Cappellaro, E., Turatto, M., Tsvetkov, D.~Y., Bartunov, O.~S., Pollas, C., Evans, R., \& Hamuy, M.\ 1997, \aap, 322, 431
\bibitem{} Cardelli, J.~A., Clayton, G.~C., \& Mathis, J.~S.\ 1989, \apj, 345, 245 
\bibitem{} Chugai, N.~N., et al.\ 2004, \mnras, 352, 1213 
\bibitem{} Chugai, N.~N., \& Chevalier, R.~A.\ 2006, \apj, 641, 1051 
\bibitem{} Davidson, K., Dufour, R.~J., Walborn, N.~R., \& Gull, T.~R.\ 1986, \apj, 305, 867 
\bibitem{} Deng, J., et al.\ 2004, \apjl, 605, L37 
\bibitem{} Deng, J., Tominaga, N., Mazzali, P.~A., Maeda, K., \& Nomoto, K.\ 2005, \apj, 624, 898
\bibitem{} Dolphin, A.~E.\ 2000, \pasp, 112, 1383 
\bibitem{} Eikenberry, S.~S., et al.\ 2004, \apj, 616, 506 
\bibitem{} Figer, D.~F., Najarro, F., Morris, M., McLean, I.~S., Geballe, T.~R., Ghez, A.~M., \& Langer, N.\ 1998, \apj, 506, 384 
\bibitem{} Filippenko, A. V. 2005, in The Fate of the Most Massive
    Stars, ed. R. Humphreys and K. Stanek (San Francisco: ASP), 34
\bibitem{} Filippenko, A. V., Li, W., Treffers, R. R., \& Modjaz, M. 2001, in 
    Small-Telescope Astronomy on Global Scales, ed. W.-P. Chen, C. Lemme, \&
    B. Paczy\'{n}ski (San Francisco: ASP), 121
\bibitem{} Gal-Yam, A., et al.\ 2004a, \apjl, 609, L59 
\bibitem{} Gal-Yam, A., Cenko, S.~B., Fox, D.~W., Leonard, D.~C., Moon, D.-S., Sand, D.~J., \& Soderberg, A.~M.\ 2004b, Bulletin of the American Astronomical Society, 36, 1408 
\bibitem{} Gal-Yam, A., et al.\ 2005a, \apjl, 630, L29 
\bibitem{} Gal-Yam, A., et al.\ 2005b, American Astronomical Society Meeting Abstracts, 207, \#78.01 
\bibitem{} Germany, L.~M., Reiss, D.~J., Sadler, E.~M., Schmidt, B.~P., \& Stubbs, C.~W.\ 2000, \apj, 533, 320
\bibitem{} Hamuy, M., et al.\ 2003, \nat, 424, 651 
\bibitem{} Hendry, M.~A., et al.\ 2006, \mnras, 529
\bibitem{} Horne, K.\ 1986, \pasp, 98, 609 
\bibitem{} Huchra, J.~P., Vogeley, M.~S., \& Geller, M.~J.\ 1999, \apjs, 121, 287 
\bibitem{} Humphreys, R.~M., \& Davidson, K.\ 1994, \pasp, 106, 1025 
\bibitem{} Iben, I., \& Renzini, A., \ 1983, ARA\&A, 21, 271 
\bibitem{} Krist, J.\ 1995, ASP Conf.~Ser.~ 77: Astronomical Data Analysis Software and Systems IV, 77, 349 
\bibitem{} Leonard, D.~C., et al.\ 2002, \aj, 124, 2490 
\bibitem{} Leonard, D.~C., Kanbur, S.~M., Ngeow, C.~C., \& Tanvir, N.~R.\ 2003, \apj, 594, 247 
bibitem{} Li, W., et al. 2000, in Cosmic Explosions, ed. S. S. Holt \& 
  W. W. Zhang (New York: AIP), 103
\bibitem{} Li, W., Van Dyk, S.~D., Filippenko, A.~V., \& Cuillandre, J.-C.\ 2005, \pasp, 117, 121 
\bibitem{} Li, W., Van Dyk, S.~D., Filippenko, A.~V., Cuillandre, J.-C., Jha, S., Bloom, J.~S., Riess, A.~G., 
\& Livio, M.\ 2006, \apj, 641, 1060 
\bibitem{} Maeder, A., \& Conti, P.~S.\ 1994, \araa, 32, 227 
\bibitem{} Matheson, T., Filippenko, A.~V., Ho, L.~C., Barth, A.~J., \& Leonard, D.~C.\ 2000, \aj, 120, 1499
\bibitem{} Maund, J.~R., Smartt, S.~J., Kudritzki, R.~P., Podsiadlowski, P., \& Gilmore, G.~F.\ 2004, \nat, 
427, 129 
\bibitem{} Maund, J.~R., Smartt, S.~J., \& Danziger, I.~J.\ 2005a, \mnras, 364, L33 
\bibitem{} Maund, J.~R., Smartt, S.~J., \& Schweizer, F.\ 2005b, \apjl, 630, L33 
\bibitem{} Maund, J.~R., et al.\ 2006, \mnras, 369, 390 
\bibitem{} Mazzali, P.~A., et al.\ 2006, ArXiv Astrophysics e-prints, arXiv:astro-ph/0603516 
\bibitem{} McCumber, M.~P., Garnett, D.~R., \& Dufour, R.~J.\ 2005, \aj, 130, 1083 
\bibitem{} Nilson, P.\ 1973, the Uppsala General Catalogue of Galaxies, Nova Acta Regiae Soc.~Sci.~Upsaliensis Ser.~V, 0 
\bibitem{} Nugent, P., et al.\ 2006, \apj, 645, 841 
\bibitem{} Oke, J.~B., et al.\ 1995, \pasp, 107, 375 
\bibitem{} Puckett, T., George, D., Ceravolo, P., Nakano, S., Sano, Y., Kushida, Y., \& Kushida, R.\ 2005, 
\iaucirc, 8615, 1 
\bibitem{} Prieto, J., Garnavich, P., Depoy, D., Marshall, J., Eastman, J., \& Frank, S.\ 2005, \iaucirc, 8633, 1 
\bibitem{} Rigon, L., et al.\ 2003, \mnras, 340, 191
\bibitem{} Ryder, S.~D., Sadler, E.~M., Subrahmanyan, R., Weiler, K.~W., Panagia, N., \& Stockdale, C.\ 2004, \mnras, 349, 1093 
\bibitem{} Ryder, S.~D., Murrowood, C.~E., \& Stathakis, R.~A.\ 2006, \mnras, 369, L32 
\bibitem{} Salpeter, E.~E.\ 1955, \apj, 121, 161 
\bibitem{} Schechter, P.~L., Mateo, M., \& Saha, A.\ 1993, \pasp, 105, 1342 
\bibitem{} Schlegel, E.~M.\ 1990, \mnras, 244, 269 
\bibitem{} Schlegel, D.~J., Finkbeiner, D.~P., \& Davis, M.\ 1998, \apj, 500, 525 
\bibitem{} Smartt, S.~J., Vreeswijk, P.~M., Ramirez-Ruiz, E., Gilmore, G.~F., Meikle, W.~P.~S., 
Ferguson, A.~M.~N., \& Knapen, J.~H.\ 2002, \apjl, 572, L147 
\bibitem{} Smartt, S.~J., Maund, J.~R., Hendry, M.~A., Tout, C.~A., Gilmore, G.~F., Mattila, S., \& Benn, 
C.~R.\ 2004, Science, 303, 499 
\bibitem{} Smartt S.~J., et al.\ 2006, in Massive stars:  From PopIII and GRBs to the Milky Way, 
STScI May 2006 (Cambridge: Cam. Univ. Press)
\bibitem{} Soderberg, A.~M., Gal-Yam, A., \& Kulkarni, S.~R.\ 2004, GRB Coordinates Network, 2586, 1 
\bibitem{} Soderberg, A.~M., Chevalier, R.~A., Kulkarni, S.~R., \& Frail, D.~A.\ 2005, ArXiv 
Astrophysics e-prints, arXiv:astro-ph/0512413 
\bibitem{} Stetson, P.~B.\ 1987, \pasp, 99, 191 
\bibitem{} Stetson, P.~B.\ 1994, \pasp, 106, 250 
\bibitem{} Stockdale, C.~J., Van Dyk, S.~D., Sramek, R.~A., Weiler, K.~W., Panagia, N., Rupen, M.~P., \& Paczynski, B.\ 2004, \iaucirc, 8282, 2
\bibitem{} Tominaga, N., et al.\ 2005, \apjl, 633, L97 
\bibitem{} Turatto, M., et al.\ 2000, \apjl, 534, L57 
\bibitem{} Turatto, M., Benetti, S., \& Cappellaro, E.\ 2003, in ``From Twilight to Highlight: The Physics of 
Supernovae'', ESO ASTROPHYSICS SYMPOSIA, Eds. W. Hillebrandt \& B. Leibundgut, Springer-Verlag, p. 200 
\bibitem{} Van Dyk, S.~D., Peng, C.~Y., Barth, A.~J., \& Filippenko, A.~V.\ 1999, \aj, 118, 2331 
\bibitem{} Van Dyk, S.~D., Garnavich, P.~M., Filippenko, A.~V., H{\"o}flich, P., Kirshner, R.~P., 
Kurucz, R.~L., \& Challis, P.\ 2002, \pasp, 114, 1322 
\bibitem{} Van Dyk, S.~D., Li, W., \& Filippenko, A.~V.\ 2003a, \pasp, 115, 1 
\bibitem{} Van Dyk, S.~D., Li, W., \& Filippenko, A.~V.\ 2003b, \pasp, 115, 1289 
\bibitem{} van Dyk, S.~D.\ 2005, ASP Conf.~Ser.~332: The Fate of the Most Massive Stars, 332, 47 
\bibitem{} Wade, R.~A. \& Horne, K.\ 1988, \apj, 324, 411
\bibitem{} Wang, L., Baade, D., H{\"o}flich, P., Wheeler, J.~C., Kawabata, K., \& Nomoto, K.\ 2004, \apjl, 604, L53 
\bibitem{} White, G.~L., \& Malin, D.~F.\ 1987, \nat, 327, 36 
\bibitem{} Wizinowich, P.~L., et al.\ 2006, \pasp, 118, 297 
\bibitem{} Wood-Vasey, W.~M., Wang, L., \& Aldering, G.\ 2004, \apj, 616, 339 


\end{thebibliography}
\end{document}